\documentclass{osa-article}

\journal{oe}

\articletype{Research Article}

\usepackage{lineno}

\begin{document}

\title{Composite acousto-optical modulation}

\author{Ruijuan Liu, Yudi Ma, Lingjing Ji, Liyang Qiu, Minbiao Ji, Zhensheng Tao, Saijun Wu}

\address{Department of Physics, State Key Laboratory of Surface Physics and Key Laboratory of Micro and Nano Photonic Structures (Ministry of Education),Fudan University, Shanghai 200433, China.}
\email{saijunwu@fudan.edu.cn}


\begin{abstract}
We propose a composite acousto-optical modulation (AOM) scheme for wide-band, efficient modulation of CW and pulsed lasers. We show that by adjusting the amplitudes and phases of weakly-driven daughter AOMs,
diffraction beyond the Bragg condition can be achieved with exceptional efficiencies. Furthermore, by imaging pairs of AOMs with opposite directions of sound-wave propagation, high contrast switching of output orders can be achieved at the driving radio frequency (rf) limit, thereby enabling efficient bidirectional routing of a synchronized mode-locked laser.  Here we demonstrate a simplest example of such scheme with a double-AOM setup for efficient diffraction across an octave of rf bandwidth, and for routing a mode-locked pulse train with up to $f_{\rm rep}=400$~MHz repetition rate. We discuss extension of the composite scheme toward multi-path routing and time-domain multiplexing, so as to individually shape each pulses of ultrafast lasers for novel quantum control applications.
\end{abstract}


%

\section{Introduction}
The developments of ultrafast technologies~\cite{Haus2000, Cundiff2003, Krausz2009,Diddams2020} have dramatically enhanced our ability to access physics at various time scales with finest precision. To utilize the pulsed lasers for nonlinear optical control~\cite{Laarmann2010,Mizrahi2013,Koch2019} and spectroscopic measurements~\cite{Maiuri2019,Picque2019}, pulse shaping techniques have been developed in the frequency and time domain to produce optical waveforms with optimally tailored spectra-phases~\cite{Weiner2000,Monmayrant2010}. Notably, existing pulse shaping techniques are designed to modulate a pulse train in a quasi-static manner. To individually modulate each output pulses from a mode-locked laser is a largely unexplored scenario, whose realization may greatly expand the applications of ultrafast lasers for quantum control~\cite{Mizrahi2013,Koch2019,He2020a,Torrontegui2020a}, metrology~\cite{Sanner2018, Yudin2020}, and imaging~\cite{Zhang2013,Andreana2015,He2017}. Unfortunately, the repetition rates of compact mode-locked lasers are high. To individually control every pulses requires precise modulation within nanoseconds, a demanding requirement in the presence of a general technical gap today on wideband laser modulation at the $\sim$GHz-level~\cite{Gould2015,Ma2020}. To modulate individual pulses, the standard method for pulse picking is through electro-optical modulation (EOM) with Pockels cells. However, the Pockels cells can hardly operate beyond a 10~MHz rate since it is difficult to generate the powerful high-voltage  waveforms while managing the dissipation~\cite{Chiow2007}. Acousto-optical modulators (AOM) transduce
low-voltage rf signals to control the optical output with accurately programmable Bragg diffraction~\cite{Donley2005,Thom2013,Zhou2020,Stummer2020}. An excellent example to illustrate AOM pulse modulation is to shift the phase of individual pulses so as to stabilize the carrier-envelop phase (CEP) of a frequency comb~\cite{Koke2010,Chen2018,Hirschman2020}. However, efficient AOM diffraction relies on phase-matching the light beams with the sound waves. Deviation of the operation parameters from the pre-aligned Bragg condition leads to reduced diffraction efficiency and distorted diffraction phases. Therefore, in traditional AOM applications, the tuning range and the control bandwidth for multi-color light are severely limited.

\begin{figure}
        \centering
        \includegraphics[width=1\textwidth]{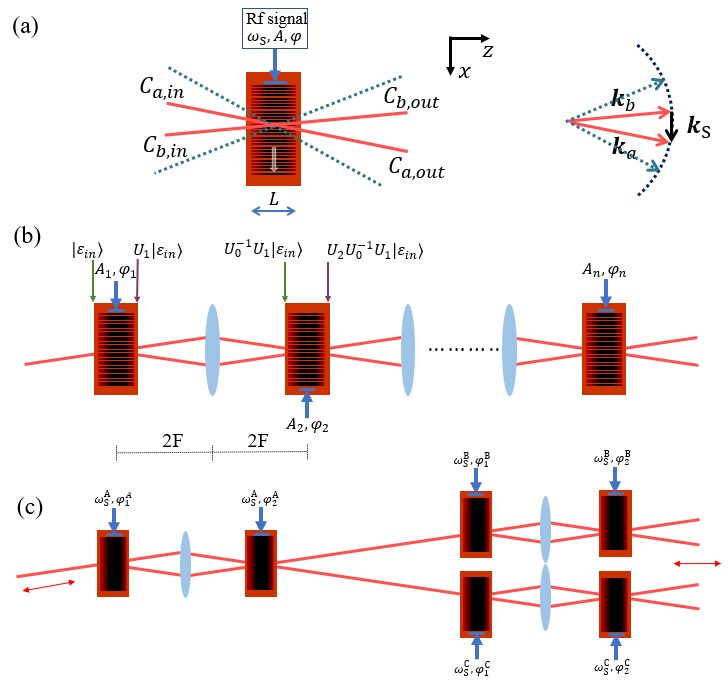}
        \caption{Operational principle of the composite AOM scheme. (a) Light beams with wavevector ${\bf k}_{a,b}$ and wavefronts $C_{a, b}$ are Bragg-coupled by the sound wave with wavevector ${\bf k}_{\rm S}$. The blue arrow points to the location of the rf-to-sound transducer represented by the horizontal blue line in the schematic, which decides the direction of sound-wave propagation (marked by the lower semi-transparent arrow). The approximate phase-matching ${\bf k}_a-{\bf k}_b\approx {\bf k}_{\rm S}$ is illustrated on the right. When the AOM is weakly driven, the diffraction orders off the Bragg condition (represented by dashed lines) can be ignored. The input-output relation for the optical wavefront can then be represented by $|\mathcal{E}_{\rm out}\rangle=U |\mathcal{E}_{\rm in}\rangle$,  
        with evolution operator $U(t,r e^{i\varphi})=(t,r e^{i\varphi};-r^* e^{-i\varphi}, t^*)$ controlled by the rf signal to transform a $(C_a,C_b)$ spinor. This input-output relation are applied iteratively in (b) for 4F-imaged identical AOMs with $U_j=U(t_j,r_j e^{i\varphi_j})$. With the output wavefront of each AOM, $|\mathcal{E}_{j,\rm out}\rangle$ imaged to the exit window of the next AOM, the wavefront at the entrance window of the next AOM is give by  $|\mathcal{E}_{j+1,\rm in}\rangle=U_0^{-1}|\mathcal{E}_{j,\rm out}\rangle$, with $U_0^{-1}$ effectively evolving the wavefront backward over the interaction length $L$ in absence of acousto-optical diffraction. (c) A simple composite AOM network with $\omega_{\rm S}^{\rm A}=\pi(2N+1) f_{\rm rep}$/2 and $\omega_{\rm S}^{\rm {B,C}}=\pi(2N'+1) f_{\rm rep}/4$ to route four temporally adjacent pulses in a $f_{\rm rep}$ pulse train from left into four output paths on the right, with close to unity efficiency and controllable relative phases. The network can be reversed for time-domain multiplexing, {\it i.e.}, to combine the relatively delayed ultrafast pulses into a single beam. 
        }\label{fig:Principle}
    \end{figure}

In this work we propose a composite AOM scheme for wide-band, efficient modulation of continuous-wave (CW) and mode-locked lasers. Inspired by the coherent control techniques for atom interferometry~\cite{Wang2005,Wu2005,Hughes2007} and nuclear magnetic resonances (NMR)~\cite{Genov2013,Genov2014,Low2016}, we identify a new class of techniques to operate AOMs beyond the Bragg condition and for rapid modulation with close-to-unity diffraction efficiency. 
The key idea is summarized in Fig.~\ref{fig:Principle}.  By optically linking a series of AOMs with 4-F imaging systems (Fig.~\ref{fig:Principle}b), the Bragg diffraction by a single AOM is coherently split into a $n$-AOM process. The $2 n-1$ amplitude and relative phase degrees of freedom, $\{A_j,\varphi_j\}$, can be continuously adjusted to optimize for specific light-control applications. 
In particular, when each ``daughter'' AOM is only weakly driven, a 2-mode approximation maps the composite diffraction dynamics to time-domain spin control, where composite pulse techniques are developed to universally enhance the resilience to the control errors~\cite{Genov2013,Genov2014,Low2016}. The key performance parameters such as the diffraction efficiency can then be optimized for {\it e.g.}, efficient diffraction of multi-color beams over a broad range of sound wave frequency even when the Bragg condition is severely violated. 

We show that an application of the composite scheme with $n=2$ already enables exceptionally high diffraction efficiency-bandwidth combinations. In particular, in this demonstration a diffraction efficiency above $85\%$ is maintained over nearly an octave of the driving rf frequency without optical re-alignments. A peak diffraction efficiency of $\sim 95\%$ is accompanied by the extinction of the undiffracted path at a $\sim 15$~dB level, allowing the double-AOM system acting as a high contrast, bi-directional router for the first time to our knowledge. When the sound waves in AOM$_2$ is chosen to propagate in the opposite direction as those in AOM$_1$ (Fig.~\ref{fig:Principle}c, Fig.~\ref{fig:DD}), the interference of the diffraction amplitudes leads to rapid switching of the output order at twice the rf driving frequency $\omega_{\rm S}$, which is exploited for routing a pulse train with nanosecond switching time.   We discuss the possibility of constructing a high-contrast composite AOM network for shaping individual pulses of a mode-locked laser with up to GHz repetition rates.

\section{Principles}\label{Sec:Theory}

\subsection{Coherent control of light with AOM}
We consider composite AOM setup schematically illustrated in Fig.~\ref{fig:Principle}b where the light diffracted by one AOM is imaged to another via a 4-F imaging system. For each single AOM operation (Fig.~\ref{fig:Principle}a), the para-axial propagation of light in the index field $n({\bf r},t)$ along ${\bf e}_z$, with $E=\mathcal{E}e^{i \bar n k_0 z}$ obeying $(\nabla^2+n^2 k_0^2)E=0$, is characterized by a slowly varying amplitude $\mathcal{E}$ with
\begin{equation}
i\partial_z \mathcal{E}=-\frac{1}{2\bar{n} k_0}\nabla_\bot^2\mathcal{E}-\delta n k_0 \mathcal{E}.\label{equ:paraaxial}
\end{equation}
Here $\bar n$ is the refractive index of the AOM crystal, for example, $\bar n\approx 2.3$ for TeO$_2$ at near infrared. An rf-driven traveling sound wave with a strain amplitude $\eta$ leads to a rapidly varying refractive index deviation $\delta n= n-\bar n $ as 
\begin{equation}
\delta n =\eta p \frac{1-\bar{n}^2}{2\bar{n}} \cos(k_{\rm S} x- \omega_{\rm S} t+\varphi),\label{equ:index}
\end{equation}
where the constant $p$ is decided by the photo-elastic effect~\cite{Feldman1978}.

The slowly varying envelops of the density wave are described by the phase $\varphi({\bf r}-{\bf v}_{\rm S} t)$ and the amplitude $\eta({\bf r}-{\bf v}_{\rm S} t)$ functions, which are controlled by those of the driving rf-fields through rf-to-sound transducers, and propagate with the group velocity ${\bf v}_S$ along ${\bf e}_x$, with $v_{\rm S}=\omega_{\rm S}/k_{\rm S}$ for the linearly dispersive crystal. For example, in TeO$_2$ the longitudinal sound velocity is $v_{\rm S}=4260~$m/s. The detailed spatial dependence is determined by the crystal geometry and the location of the transducers. Here, we simply assume a stationary and spatially uniform $\varphi$, and a stationary amplitude following a step-wise distribution as constant $\eta$ for $0<z<L$ and zero otherwise. The treatment can be generalized to account for realistic sound wave distributions in the crystal. 

The equivalence between Eq~(\ref{equ:paraaxial}) and the two-dimensional Schr\"odinger equation allows us to introduce Dirac notation $|\mathcal{E}(z)\rangle$, with $\mathcal{E}({\bf r}_{\perp},z)=\langle {\bf r}_{\perp} |\mathcal{E}(z)\rangle$, to conveniently describe the wavefront of the optical mode-function propagating along $z$. The input-output relation for the wavefront through the $0<z<L$ interaction can then be generally expressed as $|\mathcal{E}_{\rm out}\rangle = U |\mathcal{E}_{\rm in}\rangle$ by an evolution operator 
$U=e^{-i H L}$ that integrates the interaction by the Hamiltonian
\begin{equation}
    H=-\frac{1}{2\bar{n} k_0}\nabla_{\perp}^2 + \eta p\frac{\bar n^2-1}{2 \bar{n}} k_0 \cos(k_{\rm S} x+\varphi).\label{equ:Htotal}
\end{equation}
Notice this $H$  governs the paraaxial wavefront propagation along $z$, with $t$ merely as a parameter. We have thus absorbed the $-\omega_{\rm S} t$ term in Eq.~(\ref{equ:index}) into $\varphi(t)$ for  conciseness in the following. Equation~(\ref{equ:Htotal}) can be re-written in the $k$-space as $H= H_0 +\int d^2{\bf k}_{\perp} V({\bf k}_{\perp})$, with $   H_0=\frac{1}{2\bar{n} k_0}{\bf k}_{\perp}^2 $ and
\begin{equation}
    V({\bf k}_{\perp} )=\frac{K}{2}e^{i\varphi} |{\bf k}_{\perp}+{\bf k}_{\rm S}\rangle\langle {\bf k}_{\perp}|+h.c..
    \label{equ:H}
\end{equation}
Physically, $V$ couples an ``infinite'' set of wavefronts  shifted in ${\bf k}_{\perp}$-space by multiple ${\bf k}_{\rm S}=k_{\rm S} {\bf e}_x$. The coupling constant $K=\eta p k_0(\bar n^2-1)/2 \bar n$ can be referred to as a ``spatial Rabi frequency''~\cite{Wu2005}. We restrict ourselves to the input wavefronts near a lowest order Bragg-resonance, $\langle {\bf r}_{\perp}|\mathcal{E}(z=0)\rangle=C_a({\bf r}_{\perp},0) e^{- i k_{\rm S} x/2}+C_b({\bf r}_{\perp},0)  e^{i k_{\rm S} x/2 }$ with $C_{a,b}$ confined to $|k_x|<k_{\rm S}/2$ in the $k$-space. Considering a diffraction ``pulse area'' $\Theta=K L$, a weak driving condition of
\begin{equation}
\Theta\ll \frac{k_{\rm S}^2}{\bar n k_0} L \label{equ:weak}
\end{equation}
allows us to truncate the wave propagation by Eq.~(\ref{equ:paraaxial}) in $k$ to obtain $\langle {\bf r}_{\perp}|\mathcal{E}(z)\rangle\approx C_a({\bf r}_{\perp},z) e^{- i k_{\rm S} x/2}+C_b({\bf r}_{\perp},z)  e^{i k_{\rm S} x/2 }$. The pair of Bragg-coupled wavefronts obey
\begin{equation}
    \begin{array}{l}
        i\partial_z C_a=\frac{({\bf k}_{\perp}-{\bf k}_{\rm S}/2)^2}{2\bar n k_0} C_a+\frac{\Theta}{2 L}e^{i\varphi} C_b,\\
        i\partial_z C_b=\frac{({\bf k}_{\perp}+{\bf k}_{\rm S}/2)^2}{2\bar n k_0} C_b+\frac{\Theta}{2 L}e^{- i\varphi} C_a.
    \end{array}\label{equ:spinor}
\end{equation}
in the $k-$space, {\it i.e.} with $C_{a,b}({\bf k}_{\perp},z)$ to be the 2D Fourier transform of $C_{a,b}({\bf r}_{\perp},z)$. By
parametrizing $(C_a,C_b)=(\cos(\frac{\theta}{2})e^{-i\phi/2},\sin(\frac{\theta}{2})e^{i\phi/2})$ and introducing a wavevector mismatch $\delta k=k_x k_{\rm S}/\bar n k_0$, the dynamics of the spinor  is mapped to that for a unit vector ${\bf r}=(\sin \theta \cos\phi, \sin \theta \sin\phi, \cos \theta)$ on a Bloch sphere (Fig.~\ref{fig:DD}b): 
\begin{equation}
    \frac{d}{d z}{\bf r}={\bf K}\times {\bf r}.\label{equ:Bloch}
\end{equation}
Here ${\bf K}=
(K\cos\varphi, K \sin\varphi,\delta k)$ is a ``spatial Rabi vector''. 
Clearly, for incident beam with ${\bf k}_{\perp}=\{k_x,k_y\}$, $|k_x|  \ll \frac{\bar n k_0}{k_{\rm S} L}$ is required to support a ${\bf k}_{\perp}$-insensitive $\Theta$-operation near the Bragg condition. One might try to improve the broadband performance simply by reducing the interaction length $L$. However, for a fixed $\Theta$-operation such as to Bragg-deflect the incident beam with $\Theta=\pi$, a reduced $L$ typically leads to increased AOM strength $K=\Theta/L$ to compromise the weak drive condition by Eq.~(\ref{equ:weak}), resulting in high-order diffraction losses. Indeed, commercially available AOMs are usually optimized near $\Theta=\pi$ with an interaction length $L$ that balances the broadband operation with the weak drive condition. The compromise is partially responsible for the never perfect Bragg-diffraction using commercially available AOMs.

So far, we merely reformulate the standard AOM Bragg-diffraction theory with notations of quantum mechanics. Nevertheless, they allow us to confirm the validity of the 2-mode approximation and to apply composite control theory~\cite{Wu2005,Genov2014,Low2016} for the $C_{a,b}$ control next.

\subsection{Composite AOM for error-resilient modulation}
\begin{figure}
    \centering
    \includegraphics[width=1\textwidth]{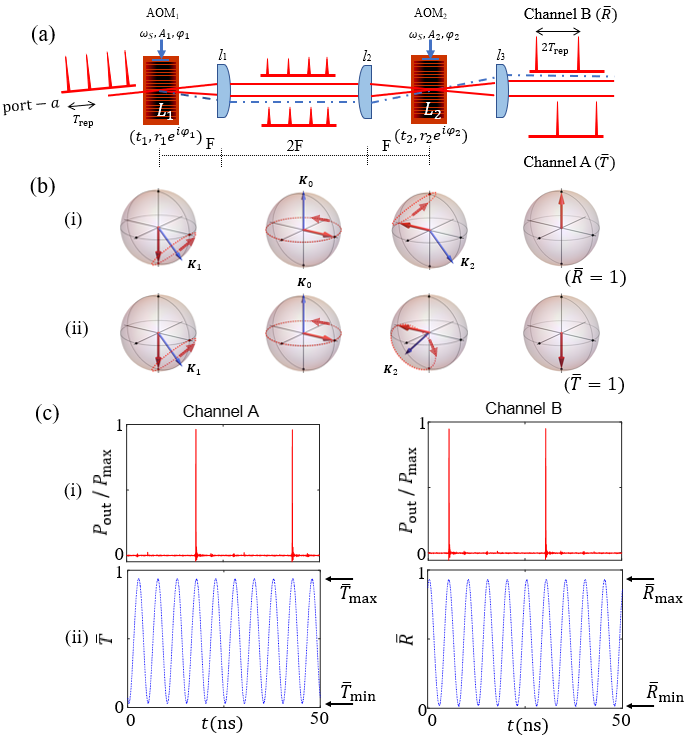}
    \caption{The double-AOM scheme for efficiently routing a mode-locked pulse train with AOM diffraction beyond the Bragg condition. In (a), AOM$_{1,2}$ with $L_1\approx L_2$ are imaged by a 4-F system, with directions of sound wave propagation effectively to be opposite to each other.
    . A pulse train from a picosecond oscillator with $T_{\rm rep}=1/f_{\rm rep}$ is subsequently diffracted by AOM$_{1,2}$. When the sound-wave frequency is set as $\omega_{\rm S}=\pi f_{\rm rep}(2 N +1)/2 $ with $N$ as an integer, $\Delta \varphi_{1,2}(t)=2\omega_{\rm S} t+\Delta \varphi_{1,2}(0)$ is incremented by $\pi$ for any pulse pair separated by $T_{\rm rep}$. By adjusting $\Delta \varphi_{1,2}(0)$ and AOM driving amplitudes $A_{1,2}$, full switching from $\bar R=1$ to $\bar T$=1 can be achieved. The effective diffraction mechanisms under the 2-mode approximation are illustrated in (b,i) and (b,ii) on a Bloch sphere, for the case of a large wavevector mismatch $\delta k\approx \pi/L\sqrt{2}$ (corresponding to the diffraction path marked with dashed blue line in (a)). Typical experimental measurements for routing a $f_{\rm rep}=80$~MHz picosecond pulses are given on (c,i) leading to 30:1 and 100:1 side-pulse suppression. The time-dependent diffraction $\bar R$ and transmission $\bar T$ measured by a CW laser are given in (c,ii), with $\{(\bar T)_{\rm max},(\bar R)_{\rm max}\}=\{0.94,0.93\}$, $\{(\bar T)_{\rm min},(\bar R)_{\rm min}\}=\{0.04,0.02\}$. The slightly decreased contrast in the CW case is likely related to the finite response time of the photo-detector. }
\label{fig:DD}
\end{figure}

We consider the setup in Fig.~\ref{fig:Principle}b where multiple AOMs are optically linked by the 4-F optics. Within the 2-mode approximation, each ``daughter-AOM''
$j$ transforms $\{C_a,C_b\}$ with a unitary matrix $U(t_j,r e^{i\varphi_j})=(t_j,r_j e^{i\varphi_j};-r^*_j e^{-i\varphi_j},t_j^*)$, which can be obtained by integrating Eq.~(\ref{equ:spinor}) and being visualized on the Bloch sphere (Fig.~\ref{fig:DD}b). Furthermore, before the second diffraction, since the 4-F system effectively images the output at $z=L$ to the exit port of the next AOM, the wavefront  should first evolve backward along $z$ with $U_0^{-1}=e^{i H_0 L}$. Here $U_0^{-1}=e^{i H_0 L}$ can also be obtained from Eq.~(\ref{equ:spinor}) with $\Theta=0$ and visualized on the Bloch sphere as precession of ${\bf r}$ around ${\bf K}_0=(0,0,-\delta k)$. More generally, the transformation of the wavefront by the composite AOM is described as $|\mathcal{E}_{\rm out}\rangle=\bar U |\mathcal{E}_{\rm in}\rangle$, with (with $\prod_{j=1}^{n}$ for multiplying from left)
\begin{equation}
\bar U=U_0 \prod_{j=1}^{n} \left( U_0^{-1}U_j\right ).\label{equ:composite}
\end{equation}
One should thus be able to adjust $\{A_j,\varphi_j\}$ of each daughter AOM to optimize $\bar U$ and the output wavefront $|\mathcal{E}_{\rm out}\rangle$. Specifically, within the 2-mode approximation, the matrix $\bar U=(\bar t,\bar r;-\bar r^*,t^*)$ is constructed in essentially the same way as those for achieving  2-level robust population inversion~\cite{Wu2005,Genov2014} and universal qubit gates~\cite{Low2016}. The wavevector mismatch $\delta k$ and diffraction pulse area $\Theta_j$ are in direct analogy to the detuning and the pulse area in 2-level spin-controls~\cite{Genov2014,Low2016}. The associated formula can thus be applied here to improve the robustness of beam-splitting~\cite{Low2016} or deflection~\cite{Wu2005,Genov2014} against deviation of these AOM operation parameters, which may arise during {\it e.g.} broadband diffraction of ultrafast pulses~\cite{Koke2010,Chen2018,Hirschman2020}. In addition, high-order diffraction leakage which generally limits the efficiency of single AOMs may be dynamically suppressed in the composite scheme, similar to those in multi-level quantum control~\cite{Genov2013}. To this end, we note similar topic to be investigated further for robust coherent control of matterwave~\cite{Wu2005,Hughes2007}.

\subsection{A double-AOM scheme for coherent optical routing}

In the following we study a simplest example of a double-AOM scheme in Fig.~\ref{fig:DD}, which is a building block of the Fig.~\ref{fig:Principle}c network, to demonstrate the utility of the composite diffraction for broadband, flexible control of CW and mode-locked lasers.

With Eqs.~(\ref{equ:spinor})(\ref{equ:Bloch})(\ref{equ:composite}), it is straightforward to understand the operation principle of the double-AOM scheme in Fig.~\ref{fig:DD}. As illustrated in Fig.~\ref{fig:DD}a, practically the 4-F optics to image the wavefronts through AOM$_{1,2}$ are composed of $l_{1,2}$ in the standard 4-F configuration to ensure the imaging precision. The sound waves in AOM$_2$ can be chosen to propagate in the same or opposite direction as those in AOM$_1$. As by Eqs.~(\ref{equ:index})(\ref{equ:H}), a flip of ${\bf k}_{\rm S}$ sign effectively change the time-dependence of the sound-wave phase $\varphi_j(t)=\varphi_j(0)\pm \omega_{\rm S} t$. 



To understand the dynamics of double-AOM diffraction, it suffices to analyze the composite $\bar U$ by Eq.~(\ref{equ:composite}) at a specific time $t$ for AOM$_{1,2}$ with a specific driven amplitude  $A_{1,2}$ and phase $\varphi_{1,2}$. According to Eq.~(\ref{equ:Bloch}), to achieve an optimal diffraction $\bar R\equiv|\bar r|^2$ (Fig.~\ref{fig:DD}(b,i)), the coherent control strategy should be that AOM$_1$ rotate the state vector ${\bf r}$ to the equator of the Bloch sphere with ${\bf K}_1$, which, after a free precession by ${\bf K}_0$, be rotated again by ${\bf K}_2$ with optimal $\Delta\varphi_{2,1}=\varphi_2-\varphi_1$ to complete the spin inversion.
Optimal diffraction with $\bar R\approx 1$ could thus be obtained even at large wavevector mismatch $\delta k\leq \pi/\sqrt{2} L$, by properly adjusting $\{A_{1,2},\varphi_{1,2}\}$. The strategy is in direct analogy to the double-diffraction method in atom interferometry~\cite{Wang2005, Wu2005, Hughes2007}. To minimize $\bar R$ so as to recover $\bar T\equiv |\bar t|^2$ to unity (Fig.~\ref{fig:DD}(b,ii)), an additional $\pi$ should be 
added to $\Delta\varphi_{2,1}$ so that AOM$_2$ undoes the rotation of ${\bf r}$ by AOM$_1$. On the other hand, the optical phase of the deflected beam can be flexibly controlled via
$\bar\varphi_{1,2}=(\varphi_{1}+\varphi_{2})/2$, such as for CEP adjustments~\cite{Koke2010,Chen2018,Hirschman2020} or coherent pulse stacking~\cite{Tunnermann2017,foot:CS}.

Importantly, in contrast to single-AOM diffraction which requires  $\Theta\approx \pi$ to achieve the best efficiency, here for the double-AOM scheme $\Theta_{1,2}=\pi/2$ are halved at $\delta k=0$. The weak drive condition by Eq.~(\ref{equ:weak}) is accordingly better satisfied toward the 2-mode approximation (Eq.~(\ref{equ:spinor})). In addition, the $U_0^{-1}$ reversal by the 4-F imaging partially rephases the $\delta k$-induced broadening, much like a spin echo in NMR, thereby improving the tolerance of efficient diffraction to the $\delta k$-spreading. These advantages are expected to improve further with larger $n$ and additional AOMs. In particular, we expect it becomes easier for the double-AOM scheme to achieve a close-to-unity composite $\bar R$ than those with single AOMs. Furthermore, when $\bar R$ is close enough to unity, the undiffracted path (Channel~A) is shut off. Therefore, by switching on/off the composite modulation, the double-AOM system should be able to route the incident light into the channel A/B, respectively. We refer this highly efficient AOM function as bidirectional optical routing. 

We now discuss the time-dependence of the double-AOM in the Fig.~\ref{fig:DD}a setup. Since the 4-F linked AOM$_{1,2}$ have opposite directions of sound-wave propagation, the relative phase $\Delta \varphi_{1,2}(t)=2\omega_{\rm S} t+\Delta\varphi_{1,2}(0)$ scans rapidly at twice the rf frequency. For the optimally adjusted $\{A_1, A_2\}$ as discussed above, an $\omega_{\rm S}=(2 N +1) \pi f_{\rm rep}/2$ with integer $N$ should result in a $\pi$ phase increment in $\Delta \varphi_{1,2}$ for two successive pulses separated by $T_{\rm rep}=1/f_{\rm rep}$. By pre-adjusting $\varphi_{1,2}(0)$, rapid switching of the composite AOM output between $\bar R=1$ and $\bar T=1$ can be achieved (Fig.~\ref{fig:DD}c) to route adjacent pulses from a synchronized mode-locked laser into two directions.  Meanwhile, the common phase $\bar\varphi_{1,2}=(\varphi_{1}+\varphi_{2})/2$ remains as a rapidly adjustable degree of freedom to shift the optical phase in the diffraction path B.



\section{Experimental demonstration}\label{Sec:Exp}

\subsection{Alignment procedure}
We demonstrate the double-AOM scheme for efficiently modulating pulsed lasers. As being summarized by Fig.~\ref{fig:DD}a, a pair of AOMs (AA optoelectronic, MT-110-A1 with an estimated $L\approx 7\sim 8$~mm)  are aligned to operate at $\omega_{\rm S}=2\pi\times 100$~MHz to Bragg-deflect the laser beam from a single-mode optical fiber into port-$a$. Phase-stable rf signals from a programmable multi-channel synthesizer (NovaTech 409B) are amplified to drive the sound waves. AOM$_2$ is imagined to AOM$_1$ by a pair of achromatic lenses $l_{1,2}$ with $F$=100~mm focal length. Here, the primary reason for choosing the achromatic lenses is to ensure precise 4F-imaging with a spatial bandwidth beyond ${\bf k}_{\rm S}$ and a field of view to cover the incident beam size. The waist radius $w\approx 110~\mu$m for the incident Gaussian beam is large enough to support $|r_{1,2}|^2\approx 80\%$ with each of the two AOMs (Fig.~\ref{fig:BB}(a)). To achieve composite AOM, the relative alignment starts with switching on a single AOM$_j$ only, with $A_j$ adjusted to achieve $|r_j|^2\approx 0.5$ half deflection. Nearly identical beam splitting by AOM$_{1,2}$ into the output Channels $A$ and $B$, with nearly identical output shapes in each case, are confirmed by recording the outputs with a digital camera. Next, fine alignments are achieved by switching on both AOM$_{1, 2}$ with reduced $|r_{1,2}|^2\sim 0.3 $ and by injecting the single-mode fiber with a synchronized pulse train with $\tau=0.5$~ns pulse duration~\cite{He2020a} at $f_{\rm rep}=100$~MHz. By shifting the $l_{1,2}$ location and finely orienting AOM$_2$, the AOM$_{1,2}$ diffractions are mode-matched, leading to uniformly varying output beams on the camera when $\Delta \varphi_{1,2}(0)$ is slowly scanned. Further improvements to the alignment is achieved by optimizing the contrast of the time-dependent transmission $\bar T$ (from Channel A) and diffraction $\bar R$ (from Channel B) at 200~MHz with a CW laser, using a multi-mode-fiber coupled fast photodetector (Thorlabs 	
DXM12DF), see Fig.~\ref{fig:DD}c for example. Finally, the rf amplitudes $A_{1,2}$ are scanned to finely locate $|r_{1,2}|_{\rm opt}^2\approx 0.5$ that maximize the composite diffraction efficiency $\bar R_{\rm max}$. Typical  $|r_{1,2}|^2$-scan 2D data for the CW laser modulation are shown in Fig.~\ref{fig:CW}a.  The optimal $|r_{1,2}|_{\rm opt}^2$ values at $\omega_{\rm S}=2\pi\times$100~MHz are marked with white dashed lines.

The moderate $\omega_{\rm S}=2\pi\times 100~$MHz in this report was chosen to fit parallel experimental progresses in our lab. In separate setups we confirm an increased $\omega_{\rm S}$ to 200~MHz does not lead to measurable degradation of the double-AOM performance. In addition to the echo-type rephasing as in Fig.~\ref{fig:DD}b to suppress $\delta k$ dephasing (Eq.~(\ref{equ:spinor})), the efficient operation of the double-AOM scheme at high $\omega_{\rm S}$ is also likely facilitated by a weaker requirement on the driving rf power than in traditional AOM applications.

\subsection{An $f_{\rm R}$=400~MHz synchronized pulse router}\label{sec:double}
From the CW laser measurements as in Fig.~\ref{fig:DD}(c,ii), we see that the full switching from $\bar R\approx 1$ to $\bar T\approx 1$ for the double-AOM output is within a $\tau_{\rm R}=2.5$~ns transient time. Therefore, even at this moderate $\omega_{\rm S}=2\pi\times 100$~MHz we are still able to realize a pulse router with nanosecond switching speed with a switching rate of $f_{\rm R}=400~$MHz. More generally, the double-AOM system can route a synchronized mode-locked pulse train with $f_{\rm rep}=f_{\rm R}/(2N+1)$ repetition rate into two directions with integer $N$. In this work, we choose $N=2$ to route a picosecond mode-locked laser (Spectra-Physics Tsunami system) with $f_{\rm rep}$=80~MHz.

To achieve the pulse routing, the Novatech rf source is synchronized with the laser pulses by phase-locking the Novatech clock with a 10~MHz laser synchronization signal. By adjusting $\Delta\varphi_{1,2}(0)$, adjacent pulses injected to port-$a$ with $T_{\rm rep}=12.5$~ns inter-pulse spacing are alternating routed into Channel A and B, as demonstrated in Fig.~(\ref{fig:DD}c).  Excluding the $\sim 5\%$ overall insertion loss, the efficiency for pulse picking in Channel A(B) reaches $96\%$ ($94\%$), with a contrast reaching 30:1 (100:1) respectively relative to the suppressed side pulses.  Notice this contrast is  compromised slightly by an $\delta \tau \approx 100~$ps relative jitter between the laser pulses and the rf signal, which is primarily limited by the phase locking quality and should be improvable to 1~ps level via {\it e.g.}, direct digital synthesizing of the rf signals.

We find $\Delta \varphi_{1,2}$ stays optimal in our temperature stabilized lab for tens of hours at least without noticeable drifts. The excellent passive phase-stability is due to common-mode rejection of drifting and vibrational noises by the two diffraction orders (Fig.~\ref{fig:DD}a), which share all the optics during their propagation and are within one millimeter to each other spatially.

We think the $\sim 95\%$ composite diffraction efficiency and the $15\sim 20$~dB side-pulse suppression are fairly close to the theoretical limit decided by the incident beam size $w$ and the interaction length $L$ (Sec.~\ref{sec:bb}). Our numerical model suggests that both the efficiency and contrast can be improved by increasing $w$ and $L$, {\it i.e.}, optimally using the apertures of large AOMs. For example, with $w=0.5~$mm and $L=15$~mm, composite efficiency $\sim 99\%$ appears achievable, a prediction subjected to experimental verification where much finer alignments are likely required particularly by taking into account variation of the sound fields experienced by the large wavefronts. A more straightforward method to further improve the performance at the moderate $w$ and $L$ is to increase the number of AOMs involved in the composite diffraction so as to further improve the small-$\Theta$ condition and to reduce the $\delta k$-broadening (Eqs.~(\ref{equ:weak})(\ref{equ:spinor})). Furthermore, the side-pulse suppression can be doubled to $30\sim 40$~dB level simply by applying the double-AOM scheme again to a selected output channel, at a small cost of overall efficiency.

\subsection{Ultra-broadband diffraction}\label{sec:bb}

\begin{figure}
    \centering
    \includegraphics[width=1\textwidth]{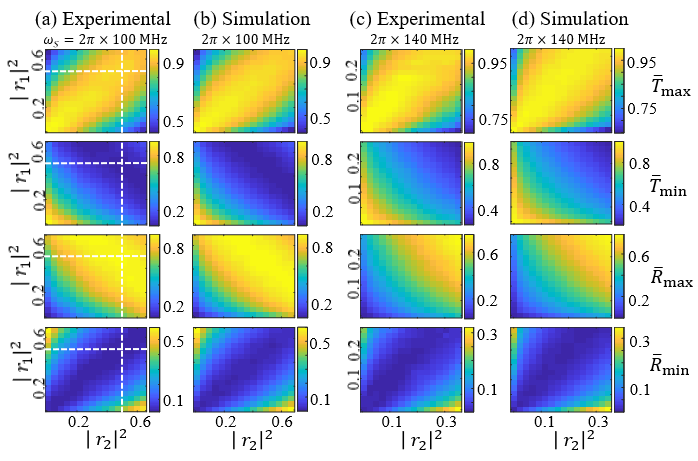}
    \caption{Fine tuning of $|r_{1,2}|^2$ to optimize $\bar R_{\max}$ for the double-AOM scheme. The experimentally measured $\bar T_{\rm max,min}$ and $\bar R_{\rm max,min}$ are given in column (a,c) (with raw data re-interpolated to evenly spaced $|r_{1,2}|^2$ grids) to be compared with the simulated data in column (b,d) based on the Eq.~(\ref{equ:paraaxial}) model with $L=8~$mm. The AOMs as by Fig.~\ref{fig:DD}a are optimally aligned at $\omega_{\rm S}=2\pi\times 100$~MHz, with corresponding data displayed in column (a,b). Data for $\omega_{\rm S}=2\pi\times 140$~MHz operation which substantially violates the Bragg condition are shown in column (c,d).}
\label{fig:CW}
\end{figure}

    \begin{figure}
        \centering
        \includegraphics[width=1\textwidth]{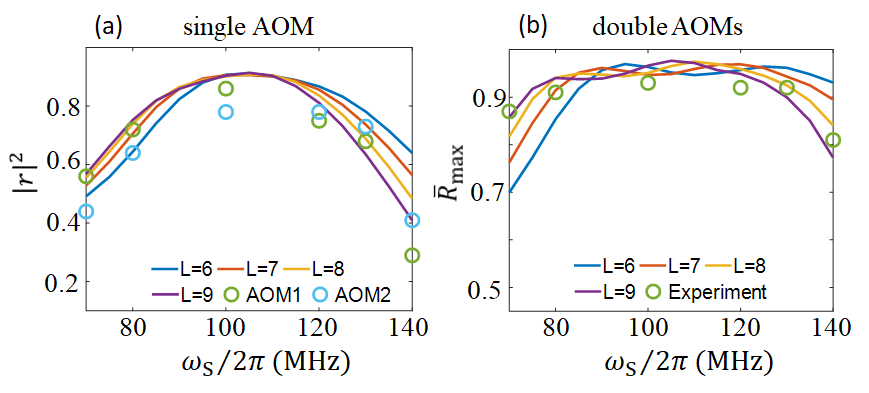}
        \caption{Optimal diffraction efficiency for single AOM (a) and double AOMs (b), measured by monitoring the CW transmission with a fast photo-detector. Simulated data based on the effective Eq.~(\ref{equ:paraaxial}) model parametrized by the interaction length parameter $L$ (mm) are displayed for comparison. For the $\bar R_{\rm max}$ measurements in (b), the efficiencies are likely to be slightly underestimated due to the finite detector bandwidth.}
    \label{fig:BB}
    \end{figure}
    
We now demonstrate the efficient double-AOM deflection  when the Bragg condition is substantially violated by a large $\delta k$ mismatch. For the purpose, we shift the sound-wave frequency $\omega_{\rm S}$ away from 100~MHz and measure the maximum diffraction efficiency $\bar R_{\rm max}$ optimized with $\{A_j,\varphi_j\}$, at sound wave frequency as large as $\omega_{\rm S}=2\pi\times 140$~MHz.  The electronic optimization process to obtain $\bar R_{\rm max}$ in Fig.~\ref{fig:CW}c is identical to those for achieving $|r_{1,2}|^2_{\rm opt}$ in Fig.~\ref{fig:CW}a for the case of $\omega_{\rm S}=2\pi\times 100$~MHz, by retrieving $\{\bar R_{\rm min},\bar R_{\rm max}\}$,$\{\bar T_{\rm max},\bar T_{\rm min}\}$ from Channel A/B with CW laser measurements (Fig.~\ref{fig:DD}(c,ii)). With the Bragg condition aligned at $\omega_{\rm S}=2\pi\times$100~MHz, a $40\%$ deviation of $\omega_{\rm S}$ leads to $k_x=0.4 k_{\rm S}$ and $\delta k=0.4 k_{\rm S}^2/\bar n k_0\approx \pi/(\sqrt{2} \times 10~{\rm mm})$. According to the analysis in Sec~3.1, if the sound field interaction length $L$ is within 10~mm, then the 2-mode theory predicts a maximum $|r_{1,2}|^2\geq 0.5$ and nearly perfect $\bar R_{\rm max}\approx 1$. Practically, we find $|r|^2_{1,2}$ saturate to 0.3 and 0.4 respectively (Fig.~\ref{fig:BB}a), likely due to a lack of rf power in our setup to fully drive the sound wave in presence of significant impedance mismatch. Nevertheless, guided by the effective theory we are able to locate $\bar R_{\rm max}=0.82$ at optimal $|r_{1,2}|^2$ and $\Delta \varphi_{1,2}$, as in Fig.~\ref{fig:CW}c and  Fig.~\ref{fig:BB}b. We further measure $\bar R_{\rm max}$ for $\omega_{\rm S}$ between 70~MHz and 140~MHz, with $\bar R_{\rm max}>90\%$ uniformly achieved between $80$ and $130$~MHz (Fig.~\ref{fig:BB}b). 



To further understand the double-AOM system beyond the 2-mode approximation, we numerically integrate the 1D AOM model outlined in Sec.~2 according to the original Eq.~(\ref{equ:paraaxial}). As discussed there, in the simple model the sound fields have a ``step-function'' distribution with an effective interaction length $L$. We simulate the performance of single and double-AOM with the model with various $L$ to match the experiment. By comparing the simulations with the experimental measurements, as in Fig.~\ref{fig:BB}, an effective $L=7\sim 8$~mm interaction length can be inferred with which we further simulate the $|r_{1,2}|^2$-tuning process as those in Fig.~\ref{fig:CW}(b,d), with very good global agreements. It should be noted that the actual sound field distribution is more complicated and may vary among AOMs. We expect even better performance once these details are included, followed by fine tuning of specific two-AOM systems.

\begin{figure}
    \centering
    \includegraphics[width=1\textwidth]{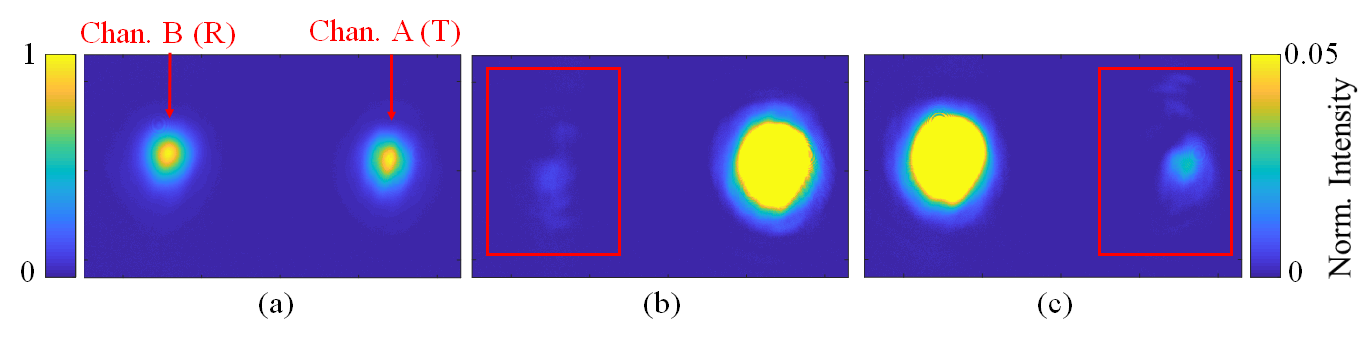}
    \caption{The output of the composite AOM recorded by a digital camera. The alignment of the double-AOM is optimized at $\omega_{\rm S}=2\pi\times 100~$MHz to route a $f_{\rm rep}=80$~MHz picosecond laser alternating into Channel A/B respectively. Here we instead set $\omega_{\rm S}=2\pi\times 80$~MHz to synchronize with $f_{\rm rep}$. By adjusting $\Delta \varphi_{1,2}(0)$, $\bar R\approx 0.5,~0,~1$ are achieved for all the pulses for the slow camera recording, as being displayed in (a),(b),(c) respectively. The re-scaled colorbar on the right is for Figs.~(b,~c) to highlight the weak residuals in the A/B channels.}
\label{fig:Img}
\end{figure}

The broadband deflection supported by the double-AOM system allows us to synchronize the composite diffraction with the mode-locked laser at  $\omega_{\rm S}=2\pi\times 80~$MHz, without realigning the optics, to directly observe the channel A/B output intensity profiles with a slow camera.  Figure~\ref{fig:Img} gives such measurements when $\Delta\varphi_{1,2}(0)$ is adjusted to give $\bar R\approx 0.5$, 0 and 1 respectively. As in Figs.~\ref{fig:Img}b,\ref{fig:Img}c, for the case of $\bar R\approx 0,~1$, the residual output in the A/B ports are so weak, that we have to over-expose the camera to see the residual features. These residual features provide us with clue on what is limiting the pulse routing efficiency and contrast for future improvements of the double-AOM system.

We note that to achieve the ultra-broadband performance with the $n=2$ composite scheme, the composite parameter $\Delta \varphi_{1,2}$ and $|r_{1,2}|^2$ controlled by each AOM needs to be adjusted for specific $\omega_{\rm S}$ in a point-by-point basis. The control bandwidth for the required adjustments of diffraction amplitudes $r_{1,2}$ by AOM$_{1,2}$ is still limited to a few MHz level, decided by the input beam size $w$ and the associated $\delta k$ spreading~\cite{Thom2013}. As discussed in Sec.~\ref{sec:double}, the ``time-reversal'' induced by the 4-F imaging enhances the $\delta k$-tolerance and therefore supports a moderately enhanced control speed. The benefit is expected to increase with $n$, a topic associated with robust AOM control in general~\cite{Genov2014,Low2016} for future investigations. On the other hand,  the double-AOM scheme enables broadband coherent frequency sweeps by rapidly following $r_{1,2}$ with $\omega_{\rm S}$ according to a pre-characterized map. Beyond the synchronized pulse routing, the MHz-level control bandwidth is typically large enough for basic and advanced CW and pulsed AOM applications~\cite{Koke2010, Endres2016} where the ability to keep nearly perfect diffraction quality is highly preferred during long-range (coherent) frequency sweeps.

\section{Discussions}\label{Sec:Dis}

\subsection{Multi-path routing and time-domain multiplexing}\label{sec:tdm}

The high pulse picking efficiency toward $99\%$ level should support one to iteratively apply the double-AOM system for multi-directional pulse routing and time-domain multiplexing of a synchronized ultrafast laser. An example of the network is illustrated in Fig.~\ref{fig:Principle}c, where the output path number is doubled with another layer of double-AOM systems. In this simple network, when the driving rf frequencies are set as $\omega_{\rm S}^{\rm A}=\pi(2N+1)f_{\rm rep}/2$ and $\omega_{\rm S}^{{\rm B,C}}=\pi(2N'+1) f_{\rm rep}/4$ and by adjusting $\Delta \varphi_{1,2}^{\rm A,B,C}$ and optimizing $A_{1,2}^{\rm A,B,C}$, an ultrafast pulse train with $f_{\rm rep}$ repetition rate from left can be routed into four output paths on the right, with stable efficiencies close to unity. The common phase $\bar \varphi_{1,2}^{\rm A,B,C}$ can be adjusted in the feedback loops to stabilize the relative phases to the desired values among the four output paths. Depending on the applications, the pulse routing contrast can be enhanced by additional time-domain-filtering  as discussed in Sec.~\ref{sec:double}, or by expanding the double-AOM units themselves into $n>2$ composite AOM sequences (Fig.~\ref{fig:Principle}b) to further enhance the performance.  

With the pulse routing speed at the driving-rf frequency limit, the composite network as in Fig.~\ref{fig:Principle}c can easily operate with ultrafast laser pulses at a GHz repetition rate, for example by choosing $\omega_{\rm S}^{\rm A}=2\pi\times 250$~MHz and $\omega_{\rm S}^{\rm B,C}=2\pi\times 125$~MHz.
After the spatial separation of the temporally adjacent pulses, one shall individually shape each pulses~\cite{Weiner2000,Monmayrant2010} and then direct them to a target experiment~\cite{Mizrahi2013,He2020a}. Alternatively, the network can be reversed to coherently combine the individually shaped pulses back to a single spatial mode via time-domain multiplexing, thereby realizing individual shaping of the adjacent ultrafast pulses at the high repetition rate. 

To reduce the insertion losses, we note the fact that the composite AOM only needs to be weakly driven makes AOM based on crystalline quartz, which are typically more difficult to be deeply driven but are more amenable to ultrafast control with low insertion losses~\cite{Vidne2003}, a favorable choice for constructing the network. In addition, since the rapid switching between $\bar R=1$ and $\bar T=1$ as in Fig.~\ref{fig:DD} is a result of diffraction interference and does not require tightly focusing the laser beams, the high-speed pulse routing can operate with much higher power than those by traditional AOM-based pulse pickers when subjecting to a same level of optical intensity damage threshold.

\subsection{Summary and outlook}

Acousto-optical modulators are ubiquitously equipped in modern research labs across fields for modulation of amplitude, phase, frequency and propagation direction of light. In this work, we have proposed a novel approach to control light with acousto-optical modulation by coherently splitting a single AOM diffraction into an n-AOM process. We have shown that the diffraction dynamics of the weakly driven n-AOM system can be mapped to time-domain spin control where composite techniques are developed to universally enhance the resilience to control errors~\cite{Genov2013,Genov2014,Low2016}. We notice related schemes of composite-pulse-inspired error-resilient 2-mode optical control have been derived ~\cite{Genov2014a, Essaadi2018,Bulmer2020} in scenarios where a mode-truncation as in this work is not required. The discussions in this work have been restricted to isotropic acousto-optical interactions. With proper modification of the simple theory, the composite diffraction technique may also be applied to acousto-optical devices based on birefringent response of slow shear waves~\cite{Tournois1997,Endres2016}. 

Experimentally, we have demonstrated the utility of the composite scheme with a simplest $n=2$ example for achieving exceptional diffraction efficiency, ultra-broadband operation, and high-contrast, bidirectional routing of a mode-locked laser.  We note that when neither the wide-band diffraction nor the $\bar\varphi_{1,2}$ diffraction phase adjustability is important, the double-AOM scheme as in Fig.~\ref{fig:DD}a can be replaced by double-passing a single AOM through retro-reflection. This can be achieved by inserting a mirror in between AOM$_{1,2}$ so that AOM$_2$ simply becomes the mirror-image of AOM$_1$. Notably, although the resulting composite AOM looks similar to traditional application of double-pass AOMs~\cite{Donley2005}, the modulation characteristics are substantially different. The results will be given in a separate technical paper. The more compact single-AOM realization of the double-AOM system can be installed, for example, for better suppressing the side pulses during the pulse routing (Sec.~\ref{Sec:Exp}). More generally, we expect extension of the kind to $n>2$ for further enhancing the performance and functionality of the composite AOM.

We expect the composite technique to enable advanced AOM applications with both CW ~\cite{Thom2013,Endres2016,Zhou2020} and pulsed lasers~\cite{Koke2010,Chen2018,Tunnermann2017, Klenke2018,Khwaja2020}. For example, the exceptionally broad rf-tuning range may help the composite AOM scheme to stabilize the carrier-envelop phase~\cite{Koke2010,Chen2018,Hirschman2020} of a few-cycle femtosecond pulse, in combination with the efficient and stable pulse compression technique recently developed~\cite{Zhang2021}.  Our method may also extend the modulation schemes and pulse shaping techniques used for coherent Raman imaging, enabling fast frequency, amplitude and polarization modulations, to achieve spectral multiplexing and background suppression~\cite{Zhang2013,Andreana2015,He2017}. Finally, as outlined in sec.~\ref{sec:tdm}, the high speed pulse routing by the double-AOM scheme  may be extended into a composite network for generating high speed, complex pulse sequence with ultrafast lasers, for novel applications in quantum information processing~\cite{Mizrahi2013,Torrontegui2020a} and nonlinear quantum optics~\cite{He2020a}. In particular, the high speed complex pulse sequence may help to extend the composite control techniques~\cite{Genov2014,Rong2015,Low2016} in NMR to optical domain for precise quantum control of strong transitions in atoms and molecules~\cite{Ma2020,Koch2019}.

\section*{Funding information}

National Key Research Program of China
(2017YFA0304204); National Natural Science Foundation of China (12074083,61975033,1874121);  The Shanghai Municipal Science and Technology Basic Research Project (19JC1410900).

\section*{Acknowledgements}

We thank Ruochen Gao, Yuxiang Zhao, Jiangyong Hu and Xing Huang for experimental assistance and discussions. We thank Prof. Yanting Zhao and Prof. Lin Zhou for helpful discussions during the development of this project.  

\section*{Disclosures}
The authors declare no conflicts of interest.

\section*{Data availability}
Data underlying the results presented in this paper are not publicly available at this time but may be obtained from the authors upon reasonable request.




\bibliography{CPAOM}

\end{document}